\documentclass[mathleft]{ano}
\usepackage{graphicx}
\usepackage{epsf}
\usepackage{cite}
\usepackage{times}
\usepackage{amssymb}
\newcommand{\ascc} {\mbox{ASCC-2.5~}}

\usepackage{natbib}
\bibpunct{(}{)}{;}{a}{}{,}
\def\aj{AJ~}%
\def\araa{ARA\&A~}%
\def\apjs{ApJS~}%
\def\apss{Ap\&SS~}%
\def\aap{A\&A~}%
\def\aaps{A\&AS}%
\def\mnras{MNRAS~}%
\def\pasp{PASP~}%
\begin{document}


\title{Open clusters and the galactic disk}

\author{S.~R\"{o}ser\inst{1}\fnmsep\thanks{Corresponding author:
  \email{roeser@ari.uni-heidelberg.de}\newline}
\and N.V.~Kharchenko \inst{1,2,3}
\and A.E.~Piskunov \inst{1,2,4}\and
E.~Schilbach  \inst{1} \and
R.-D.~Scholz \inst{2} 
\and  H. Zinnecker\inst{2}
}
\titlerunning{Open clusters and the galactic disk}
\authorrunning{S.~R\"{o}ser et al.}
\institute{Astronomisches Rechen-Institut, Zentrum f\"ur Astronomie der
    Universit\"at Heidelberg, M\"onchhofstr. 12-14, 69120 Heidelberg,
    Germany
\and 
Astrophysikalisches Institut Potsdam, An der Sternwarte 16, 
D-14482 Potsdam, Germany
\and
Main Astronomical Observatory, 27 Academica Zabolotnogo Str., 03680
Kiev, Ukraine
\and
Institute of Astronomy of the Russian Acad. Sci., 48 Pyatnitskaya
Str., 109017 Moscow, Russia}

\received{8 April 2010}
\accepted{}
\publonline{Astronomische Nachrichten and Reviews in Modern Astronomy 22 (2010).}


\abstract{
It is textbook knowledge that open clusters are conspicuous members of the thin disk of our Galaxy,
but their role as contributors to the stellar population of the disk was regarded as minor.
Starting from a homogenous stellar sky survey, the \ascc, we revisited the population
of open clusters in the solar neighbourhood from scratch. In the course of this enterprise we detected
130 formerly unknown open clusters, constructed volume- and magnitude-limited samples of clusters,
re-determined distances, motions, sizes, ages, luminosities and masses of 650 open clusters.
We derived the present-day luminosity and mass functions of open clusters (not the stellar mass function in open clusters), the
cluster initial mass function CIMF and the formation rate of open clusters. We find that open clusters 
contributed around 40 percent to the stellar content of the disk during the history of our Galaxy.
Hence, open clusters are important building blocks of the Galactic disk.}

\maketitle

\section{Introduction}

Open clusters constitute an important part of a process transforming gas and dust into stars.
They are observed as the most prominent parts in the regions of active star formation,
or as tracers of the ceased star formation process in the general Galactic field.
However, the role they are playing in this process has still not been fully understood.
In spite of their prominence, there had been indications that classical open clusters contribute
only 10\% or even less input \citep{1971A&A....13..309W,1978PASP...90..506M,2006A&A...445..545P}
to the total stellar population of the Galactic disk. This discordance can be
explained either by an early decay of a considerable fraction of newly formed
star clusters \citep[see e.g.][]{2005A&A...441..117L,2001MNRAS.321..699K,1978A&A....70...57T}
or by an insufficient knowledge of cluster formation statistics. In this context,
one should note that the most important items of cluster formation like the distribution
of cluster masses at birth (i.e., the initial mass function of star clusters)
and the cluster formation rate were still poorly known a decade ago.

In principle, basic parameters
like distance, motion, age, and metallicity can be determined for an open cluster
more accurately than for a single field star. Indeed, they are better tracers of large scale structures
of the Galactic disk population than field stars. Nevertheless, the  
most comprehensive  studies of the Galactic cluster population are about 20
years old \citep{1982A&A...109..213L,1988AJ.....95..771J}. They were
based on the best data available at that time, the
Lund Catalogue of Open Cluster Data \citep[][hereafter, the Lund Catalogue]{lyn87}
and its subset of  clusters with 3-colour photometry \citep{1982ApJS...49..425J}.
Although these studies represent an important
step in our understanding of the general properties of the cluster population,
they suffer from  incompleteness of the cluster samples and from inhomogeneity
of the cluster parameters. 

About
1200 clusters were known in the Lund catalogue by 1988. Only 400 of them had
accurate, but heterogeneous UBV photometry, and photometric distances, reddening
and age values. Although for almost all clusters apparent diameters were given
in the Lund catalogue (estimated by eye from sky charts or defined by the size
of detector field of view), only about 100 clusters were
studied in a systematic way on the basis of star counts \citep{1994A&AT....6...85D}.

Currently, the on-line list of open cluster data by \citet[][DLAM hereafter]{2002A&A...389..871D},
which can be considered as a continuation
of the Lund Catalogue, contains by a factor of 1.5 more clusters than its
predecessor, but the degree of completeness of this list is still unknown. Since
the cluster data in the DLAM list are taken from literature,
the set of the derived parameters differ from cluster to
cluster. Also, the parameters themselves are based on heterogeneous observations and
different methods of cluster definition and of parameter determination. Whenever using
these data for cluster population studies, one is confronted with problems caused
by uncertain cluster statistics and data heterogeneity.

In the following sections we decribe our approach to construct a representative sample
of open clusters in the solar neigbourhood, derive a homogeneous set of cluster parameters,
especially age, mass and luminosity which are basic for estimating the role of open
clusters as building blocks in the evolution of the Galactic disk.
A previous review on this topic was given by \citet{2009IAUS..254..221Z}
at IAU Symposium 254.

\section{A volume- and magnitude-limited sample of open clusters}

The first goal in this cooperation which started in 2003 was the construction
of a sample of Galactic open clusters whose properties and biases are known (as good as possible).
In this approach we started from the very beginning, namely from a magnitude-limited
catalog (sky survey), the 
All-Sky Compiled Catalogue of 2.5 million stars 
(\ascc\footnote{\texttt{ftp://cdsarc.u-strasbg.fr/pub/cats/I/280B}}).
The latest version of \ascc \citep{2009yCat.1280....0K} \linebreak
gives absolute proper motions in the ICRS, Johnson
$B$, $V$ and 2MASS $J, H, K_s$ as well as spectral types and radial velocities if available.
The catalog is to 90\% complete down to V = 11.5.
\ascc was used to
identify known open clusters and compact associations from the Lund Catalogue \citep{lyn87},
the \citet{2002A&A...389..871D}
on-line data collection,
and the \citet{1981csca.book.....R} list of associations.
Cluster membership of stars was determined based on kinematic and photometric criteria.
In the \ascc we
found 520 of about 1700 known clusters \citep{2005A&A...438.1163K}, 
and discovered 130 new open clusters
\citep[][]{2005A&A...440..403K}.

\begin{figure}[h!] 
\resizebox{\hsize}{!}{
\includegraphics[bb=95 210 580 625,clip,width=0.99\hsize]{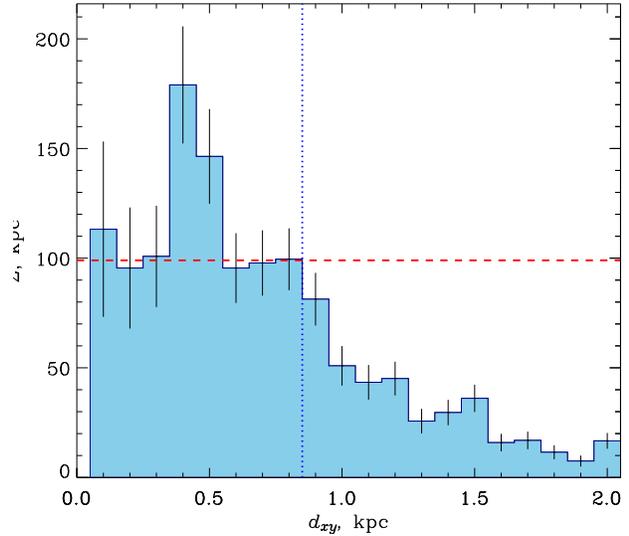}}

\caption{Distribution of the surface density $\Sigma$ of open clusters versus
distance $d_{xy}$ from the Sun projected onto the Galactic plane. The
dotted line indicate the completeness limit, the dashed horizontal line
correspond to the average density of ``field clusters''.
The peak at $d_{xy}$ = 0.4 to 0.5 is due to a population of young
clusters mainly connected to Gould's belt. The bars are Poisson errors derived from cluster counts.
From \citet{2006A&A...445..545P} where a detailed anaysis can be found. 
}\label{ddist_fig}
\end{figure}

From this full sample of 650 clusters in the \ascc \linebreak
we extract 2 important sub-samples:
a volume-limited sample of 256 open clusters complete to a distance
of 850\,pc from the Sun, see \citet{2006A&A...445..545P} and Fig.\ref{ddist_fig},
and a magnitude-limited sample
complete down to apparent integrated magnitude $I_V$ = 8, with 440 clusters  above this completeness limit.
For details of the construction of the magnitude-limit\-ed sample see \citet{2008A&A...487..557P}.
Finally, we must keep in mind the evolutionary status of open clusters included
in our sample. Since cluster membership is based on the proper motion data mainly
obtained in the optical spectral range, we consider our sample as representative
of optical clusters or ``classical'' open clusters. Embedded objects are not included in this sample since their
members usually are fainter (in the visual) than the limiting magnitude of the
\ascc. Therefore, we assume the beginning of the
transparency phase after the removal of the bulk of the placental matter to be a
starting point of the evolution of a classical open cluster. The corresponding
age $t_0$ is defined by the lowest age of our clusters, that is, about 4 Myr.

\subsection{Distribution in the Galactic Disk}\label{surf-dist}

Having reliably determined membership of stars in open clusters, the distances from the Sun together
with the interstellar extinction to the cluster was derived \citep{2006A&A...445..545P}. 
The symmetry plane of the clusters' distribution is determined to be at $Z_0=-22\pm4$ pc,
and the scale height of open clusters is only $56\pm3$ pc. The total surface density (Fig. \ref{ddist_fig})
and volume density in the symmetry plane are $\Sigma=$ 114 kpc$^{-2}$ and
$D(Z_0)=1015$ kpc$^{-3}$, respectively. We estimate the total number of open clusters in
the Galactic disk to be of order of 10$^5$ at present.

\section{Astrophysical parameters of open clusters}\label{ast-par}

\subsection{Age distribution of open clusters}

Ages of open clusters in the Galaxy can be determined from the
fitting of theoretical isochrones to the loci of member stars in the CMD.
It should be kept in mind that
this makes the basic difference to the age determination of open clusters
in other galaxies.
Our method of age determination is described in detail in \citet{2005A&A...438.1163K}.
Ages were determined for 506 out of 520 clusters, of which 196 are
first estimates. Our values for cluster ages are in good agreement
with earlier results by 
\citet{lok01} and \citet{lok04}.

\begin{figure}[h!]
\resizebox{\hsize}{!}
{\includegraphics[bb=60 200 535 670,clip,width=0.99\hsize]{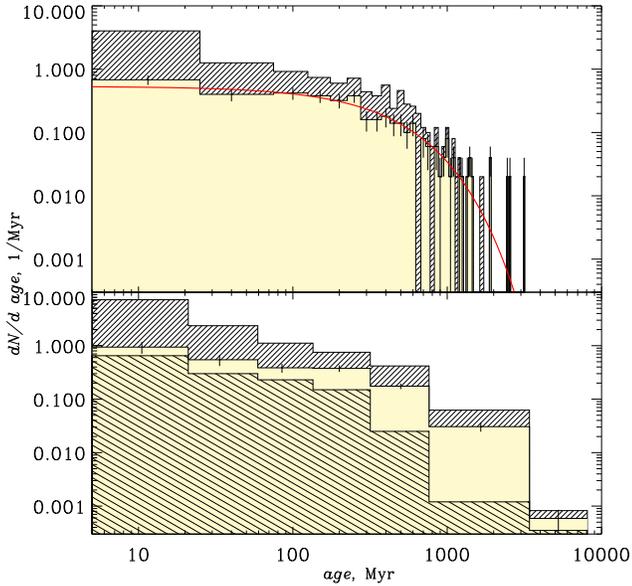}}
\caption{Distribution of open clusters versus age. For an easier comparison,
the distributions for the different samples are not normalised to unit area.
Upper panel: our total sample is shown as the hatched histogram.
The sample of field clusters within the completeness area is marked as the filled histogram,
and the solid curve marks the fitted age distribution. The age step is 50 Myr.
Lower panel: the same distributions as in the upper panel together 
with the age sample used by \citet{1971A&A....13..309W}.
The data from \citet{1971A&A....13..309W} are shown as the backhatched histogram in
the foreground. The vertical bars correspond to Poissonian errors derived from
cluster counts. The binning is chosen same as in \citet{1971A&A....13..309W}.
From \citet{2006A&A...445..545P}.
}\label{dndt_fig}
\end{figure} 

In the upper panel of Fig.~\ref{dndt_fig} the distributions of clusters versus 
age are shown for the total sample as well as for those
within the completeness area. Since young clusters contain,
in general, more luminous stars, they can be observed at larger distances
(beyond 850 pc) and their proportion in the total sample
is somewhat higher than that of older clusters. Hence
the total sample is biased towards young clusters.
Such a bias has a strong impact onto the determination of cluster formation rate
and lifetime. Therefore we determined these parameters from the
volume-limited sample.

In the bottom panel of Fig.~\ref{dndt_fig} we show the same distributions  
together with results from \citet{1971A&A....13..309W} based on the \citet{1971A&AS....4..241B} sample.
The pronounced deficiency of older clusters in the latter sample is the reason
for the smaller mean lifetime derived by \citet{1971A&A....13..309W} of 231 Myr compared
to the 327$\pm$25 Myr for our volume-limited sample.

\subsection{Integrated magnitudes and colours of open clusters}

\begin{figure}[!]
\includegraphics[angle=270,clip,width=0.99\hsize]{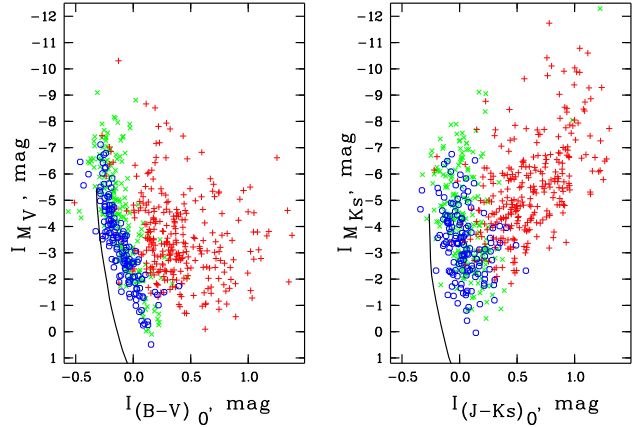}
\caption{Integrated magnitudes and colours of 650 Galactic open clusters.
Left panel: the colour-magnitude diagram $I_{M_V}$ vs. $I_{(B-V)_0}$.
Right panel: the colour-magnitude diagram $I_{M_{K_s}}$ vs. $I_{(J-K_s)_0}$.
Clusters with only MS-members are 
marked by open circles, clusters with evolved members 
which did not yet cross the Hertzsprung gap are marked 
by crosses, clusters containing red giants are marked by 
plusses. It is remarkable that in the optical most of the brightest clusters
have blue colours, whereas in the NIR they have red colours.
}
\label{fig:clustercmd}
\end{figure}

In the past decades quite a number of authors determined the integrated colours and magnitudes
of MW open clusters. Among them we mention \citet{1965AJ.....70..362G},
\citet[]{1974NInfo..33..101P}, \citet[]{1983BASI...11...44S}, \citet[]{1985Ap&SS.112..111S},
\citet[]{1989MNRAS.236..263P}, \citet[]{1994A&AS..104..379B} and \citet[]{2002A&A...388..158L}.
In many cases, however, the underlying data are strongly inhomogeneous since the
photometric observations of different clusters were obtained with different
instruments and detectors, and the data reduction was carried out with
different methods by different authors. Frequently, the integrated magnitudes
and colours were only ``by-products'' of studies aiming primarily at 
constructing photometric sequences (e.g., sets of photometric standards,
or cluster CMDs), \linebreak where the data completeness is not essential.

In Fig. \ref{fig:clustercmd} we show the colour-absolute magnitude diagrams for 650 clusters in the
solar neighbourhood.
At first sight, it looks rather surprising that, in the optical, most of the open clusters in the
solar neigbourhood appear rather blue even if they are not as young as, e.g., open clusters
in galaxies with active star formation. \citet[]{2009A&A...507L...5P} discuss this finding
and show that this blue colour must be expected in clusters with masses less than about 10$^4$~$M_\odot$
where due to the discreteness of the IMF red giants do pop up only for a short period during a cluster's
lifetime. Hence, the usually adopted SSP models are not suited to reproduce the colours of the open clusters in
the solar neighbourhood.

\subsection{The luminosity function of Galactic open clusters}

Even in the close vicinity of the Sun, the only previous attempt to construct the luminosity
function of open clusters \citep{1984AJ.....89.1822V} is based on a sample of 142
clusters that is to 2/3 complete within 400 pc.

\begin{figure}
\resizebox{\hsize}{!}{
\includegraphics{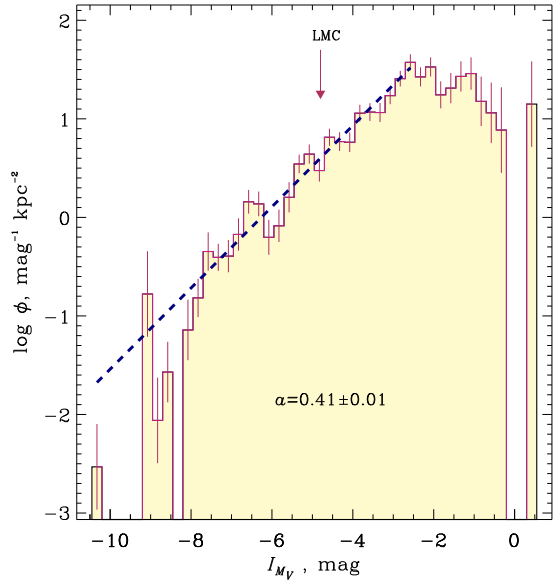}
}
\caption{Luminosity function of Galactic open clusters based on 440 local
clusters brighter than the completeness limit $I_V$ of the sample. The
bars are Poisson errors, the dashed line shows a linear fit for the brighter
part of the histogram ($I_{M_V} < -2.5$) where $a$ is the corresponding slope.
The arrow indicates the limit of integrated absolute magnitudes reached for open
clusters in the LMC \citep[see][]{2002AJ....124.1393L}. Figure from \citet{2008A&A...487..557P}.
}
\label{fig:histolf}
\end{figure}

Fig.~\ref{fig:histolf} \citep[from][]{2008A&A...487..557P}  shows the present-day luminosity function (CPDLF)
constructed from the \linebreak mag\-ni\-tude-limited sample of 440 clusters.
At brighter magnitudes the CPDLF follows a power law with an exponent  $\alpha$ in $dN/dL_V
\propto L^{-\alpha}$ which comes out as $\alpha=2.02\pm0.02$.
This is comparable to the slope in extragalactic clusters \citep[see e.g.][]{2002AJ....124.1393L}.
Notice, that for Galactic star clusters the
CPDLF can be observed much deeper than for clusters in other galaxies. For the
Large Magellanic Cloud (LMC), the faint limit is reached already at about
$I_{M_V}=-5$, and it is much brighter in more distant galaxies
\citep[][]{2002AJ....124.1393L}. As a consequence of going deeper we find 
a turnover in the CPDLF between $I_{M_V}=-3$ and $-2$, and an apparent
decrease at fainter magnitudes. This turnover is a real phenomenon,
since the luminosity function is obtained from the distribution of clusters
within the completeness limit.

\subsection{Masses and the mass function of Galactic open clusters}

Mass is one of the fundamental parameters of star clusters which,
from an observational point of view, is difficult to determine. 

There are at least three
independent methods for estimating cluster masses,
each with advantages and disadvantages.
The simplest and most straightforward way is to \linebreak count cluster members and to sum
up their masses.
This requires a complete census of cluster
members (down to the lowest masses). The real situation is, however, far
from being ideal: incompleteness comes from either the limiting magnitude
or the limited field of view or both. The extrapolation
of the observed mass spectrum to ``unseen'' cluster members by choosing
some inappropriate IMF can lead to unjustified and unpredictable 
modifications of the observed cluster mass, i.e. to biases. Nevertheless, due to its
simplicity, the method is currently widely used \citep[see][]{1994A&AT....6...85D,
2002NewA....7..553T,2005A&A...441..117L}.

\begin{figure}[!]
\centering
\includegraphics[angle=270,clip,width=0.7\hsize]{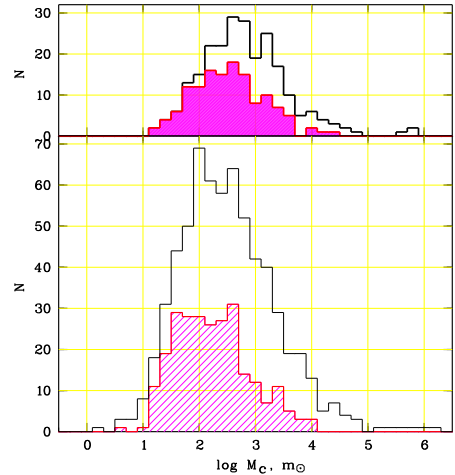}
\caption{Distributions of tidal masses of open clusters. The upper panel shows
masses calculated from measured radii for 236 clusters. The lower panel displays
the masses calculated from calibrated radii of all 650 clusters of our sample.
The open histograms correspond to total distributions, the filled/hatched ones
are distributions of corresponding clusters residing in the area where our sample
is complete. Figure from \citet{2008A&A...477..165P}.}\label{fig:mass-his}
\end{figure}

Another method is based on the virial theorem:
the mass of a cluster is determined
from  the
stellar velocity dispersion. It
does not require the observation and membership determination of all cluster
stars.
The application of the meth\-od is,
however, limited to sufficiently massive stellar systems (globular clusters and
dwarf spheroidals)
with dispersions of internal motions large enough to be measurable.
For open clusters, the typical velocity dispersion is about or less than 1~km\,s$^{-1}$,
so, only for a few clusters with the most accurate proper
motions and/or radial velocities, attempts have been undertaken to derive ''virial 
masses'' \citep[see the references in][]{2007A&A...468..151P}.

The third method uses the interpretation of the tidal interaction of a cluster
with
the parent galaxy, and requires knowledge of the tidal radius of a cluster.
It gives so-called 'tidal masses', and goes back to the fundamental paper by \citet{1962AJ.....67..471K}.
Considering globular clusters which, in general, have elliptical orbits,
\citet{1962AJ.....67..471K} differentiates between the tidal and the limiting
radius of a cluster. For open
clusters revolving at approximately circular orbits, one can expect the observed
tidal radius to be approximately equal to the limiting one, although a probable
deviation of the cluster shape from sphericity may have some impact on the
computed
cluster mass.
Nevertheless, this method gives a mass estimate of a cluster \citep{1998A&A...333..897R,1998A&A...329..101R}
that is independent
of the
results of the two methods mentioned above. As tidal masses grow with $r_{t}$ cubed, their precision
is strongly influenced
by the uncertainties of the $r_{t}$ determination. For details
of the application of King's model the reader is referred to \citet{2007A&A...468..151P}
and \citet{2008A&A...477..165P}.

For 236 clusters of our sample
we could determine
core and tidal radii directly from the fitting of King profiles to the density profile of
cluster members.
The distribution of the corresponding cluster masses is shown in the upper part of Fig.~\ref{fig:mass-his}, where
the filled histogram shows the distribution within the completeness area. Most
of the clusters have tidal masses between 50 and 5000 $m_\odot$, and
for about half of the clusters, the masses were obtained with a relative error
better than 50\%. To obtain a mass estimate for all the 650 clusters we 
calibrated the semi-major axis $\cal A$ (of the observed stellar density distribution) of the
clusters using the tidal radii of the sub-sample of 236 clusters. The resulting mass
distribution is shown in the lower part of Fig.~\ref{fig:mass-his}. Although the masses
are of moderate accuracy the large cluster sample should lead to reliable
statistical evaluation.

\begin{figure}
\resizebox{\hsize}{!}{
\includegraphics{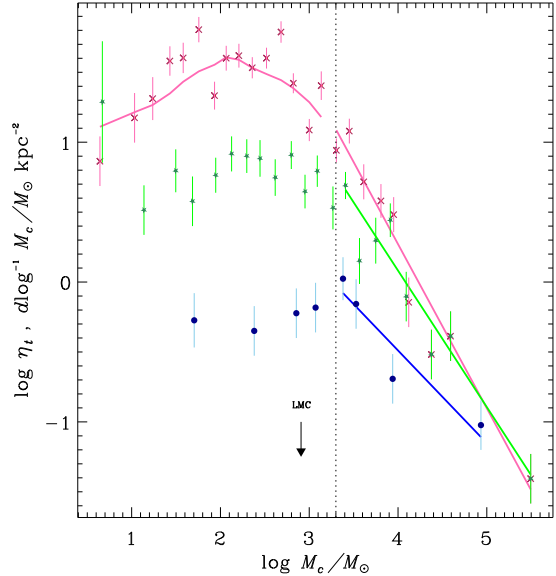}
}
\caption{Evolution of the mass function $\eta_t = {dN}/{d\log M_c}$ of Galactic open
clusters. Different symbols mark samples with different upper limits of cluster
ages. The filled circles are for clusters with $\log t \leqslant 6.9$ (CIMF), stars
show the CMF for $\log t \leqslant 7.9$, and crosses indicate the CPDMF based on all
440 clusters ($\log t \leqslant 9.5$). The bars are Poisson errors. The straight
lines are the corresponding fits to linear parts of the mass functions at masses
greater than $\log M_c=3.3$ indicated by the vertical dotted line. The curve is
the smoothed CPDMF at $\log M_c<3.3$. The arrow indicates the lower mass limit
reached for open clusters in the LMC. Figure from \citet{2009A&A...507L...5P}.
}
\label{fig:fuma}
\end{figure}

The corresponding present-day mass function CPDMF 
is shown as the upper curve in
Fig.~\ref{fig:fuma}. Notice that in Fig.~\ref{fig:fuma} we show the logarithmic mass function in the
form $\eta_t = {dN}/{d\log M_c}$ while in the further discussion (and to compare with other authors)
we refer to the mass function \linebreak $dN/dM_c$.
On its high-mass part
($\log M_c>3.3$) the \linebreak CPDMF follows a power law with an exponent  $\alpha$ in
\linebreak $dN/dM_c$
$\propto M_c^{-\alpha}$ which comes out as $\alpha=2.17\pm0.08$. \citet{2003MNRAS.343.1285D}  derived masses for open
clusters in four galaxies and found a typical value of the CPDMF slope $\alpha\approx2$
within a mass range $(10^3,10^6\,M_\odot)$. This is in remarkable agreement with the slope found for our CPDMF.
Notice, however, that the
masses in
\citet{2003MNRAS.343.1285D} are based
on the mass-luminosity relation used to convert the observed photometric or
spectroscopic data into masses. Nevertheless, this can be interpreted as indirect evidence of the coincidence of
the relative mass scales
of Galactic and extragalactic clusters.


In Fig.\ref{fig:fuma} we also show  the cluster mass function with different upper limits for age.
The youngest clusters with $\log t\leqslant6.9$ (49 in our sample) are considered to
represent the open cluster initial mass function (CIMF).
It is shown
as the lowest curve (with solid dots).
The CIMF has a power-law shape with 
$\alpha=1.66$ between $\log M_{\rm c}\approx 3.3$ and the cut-off at about $\log M_{\rm c}\approx 5$.
The low-mass part ($\log M_{\rm c}\leq 3.3$) can also be fitted by a power-law with $\alpha=0.82\pm0.14$.
With time, the slope of the high-mass part increases, for clusters with $\log t\leqslant8.5$ we find $\alpha=2.13$,
and for $\log t\leqslant9.5$, the CPDMF, $\alpha$ increases to 2.17.  
So, at every age the cluster mass function shows the 
same basic features, i.e., a quasi-linear high-mass portion,
and a non-linear portion at lower masses. The
low-mass portion changes from an approximately flat distribution at $\log t=6.9$
to a clearly non-linear behaviour displaying a broad maximum with a peak at
about 100 $M_\odot$ for the CPDMF and a decline towards lower masses.

With the same
completeness arguments as for the \linebreak CPDLF we infer that this maximum of the CPDMF and the decline towards
lower masses is real and not due to an observational bias. Moreover, the maximum in the
CIMF at about 1000~$M_\odot$ and the decline shortward is also real, as the detection
probability of young clusters of, say, a few hundred $M_\odot$ is larger than that of old
clusters, because the former contain brighter stars.

The steepening of the time-integrated mass function is a direct consequence of the
mass-loss of clusters due to dynamical evolution together with the cut-off at
the high-mass end of the CIMF. When the clusters grow older the high-mass cut-off shifts towards
lower masses. Hence, the number of clusters in the higher mass bins increases more slowly or does not
increase at all. This leads to a steepening as well as a shift of the maximum to lower masses. 

\begin{figure}[!]
\includegraphics[bb = 9 52 1286 1168,angle=0,clip,width=0.9\hsize]{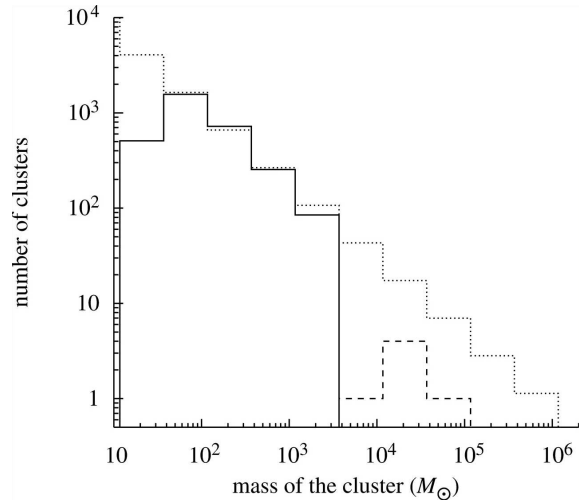}
\caption{The embedded-cluster mass function (ECMF) for the Galaxy. The scaled ECMF for clusters from \citet{2003ARA&A..41...57L}
  compilation of cluster masses within
  2.5~kpc is plotted as the solid line.
   The predicted ECMF for all masses and a spectral index
    of $\alpha = 1.7$ is shown as the dotted line.
     Massive embedded clusters from the list of \citet{ascenso08}
      are represented by the dashed line. Adapted from \citet{ascenso08}
      and \citet{2010RSPTA.368..713L}.
}
\label{fig:lada-ascenso}
\end{figure}

According to present belief, the classical gas-free open clusters stem from a population
of clusters which are surrounded by the remnants of the molecular cloud in which their
stars have formed. 
\citet{2003ARA&A..41...57L} have compiled a catalogue of about 100 embedded clusters within 2.4
kpc from the Sun. The sample contains some optical objects and is partly
overlapping with our data. Using models of the luminosity function,
\citet{2003ARA&A..41...57L} scaled the IR counts within the areas studied, estimated cluster
masses, and constructed an embedded cluster mass function (ECMF) shown in Fig. \ref{fig:lada-ascenso}. Typically, the
clusters are distributed over a mass range from 50 to 1000~$M_\odot$ and follow a power law
of the form  $dN/dM_c
\propto M_c^{-\alpha}$, where $\alpha \approx  1.7$ to 2.0.
There are striking similarities between the ECMF and the CIMF.
Both follow a power law with about the same exponent $\alpha \approx  1.7$,
which hints that both groups come from a universal parent distribution. Also, both show cut-offs.
The high-mass cut-off of the ECMF coincides remarkably well with the low-mass cut-off of the CIMF.
\citet{2003ARA&A..41...57L} also determined the embedded cluster formation rate to be
about $2-4\,\mathrm{kpc}^{-2}\mathrm{Myr}^{-1}$ which is about 10 times larger than our
open cluster formation rate of $0.4$ kpc$^{-2}$Myr$^{-1}$ determined from the CIMF in Fig.\ref{fig:fuma}.
The latter low rate led \citet{2003ARA&A..41...57L}
to the conclusion that about 10\% of the embedded clusters do survive to become
classical open clusters. Hence it is not surprising that the CIMF in Fig.\ref{fig:fuma}
has a break at about 1000~$M_\odot$. On the other hand, one could ask why
in Fig. \ref{fig:lada-ascenso} the 
embedded counterparts of the classical open cluster more massive than about 1000~$M_\odot$
are absent. In fact only a few of them have been detected in a recent work by
\citet{ascenso08}.  
An answer to this question may have already been given by \citet{2002MNRAS.336.1188K}. They
found that (initially embedded)
clusters that form in total $10^3 < N < 10^5$ stars (so-called type II clusters) lose their gas
within a dynamical time as a result of the photoionizing flux from O stars.
Sparser (type I) clusters get rid of their residual gas on a time-scale longer
or comparable to the nominal crossing time and thus evolve approximately adiabatically.
For \citet{2002MNRAS.336.1188K} this effect works on the transformation
of the mass function of embedded clusters (ECMF)
to the `initial' mass function of bound gas-free star clusters (CIMF).
They estimate that the resulting ICMF has, for a featureless power-law ECMF,
a turnover near $10^{4.5}$ $M_\odot$ and a peak near $10^3$ $M_\odot$.
This explains both the absence of high-mass clusters in the ECMF and the
low number of low mass clusters in the CIMF. The latter being related to
'infant mortality'.

\section{Evolution of open clusters and their contribution to the disk population}

The driving forces in the modification of the cluster mass function with time lies
in the evolution of the individual clusters during their life-time: mass-loss
both from stellar evolution of massive stars and from dynamical evolution affecting
preferentially low-mass stars.

This mass-loss of clusters is determined from
comparing the average mass of the newly formed, youngest clusters
${M}_c\simeq 4.5\cdot 10^3\,M_\odot$ with the average cluster mass
from the whole sample (${M}_c\simeq 700\,M_\odot$) \citep{2008A&A...487..557P}.
The typical mass-loss occurring in open clusters
during their evolution amounts to about 3-14 $M_\odot$ Myr$^{-1}$.
In the earliest phase of the cluster evolution this mass loss primarily occurs
from stellar evolution of massive stars and even from the expulsion of massive
stars from the cluster. \citet{2008A&A...489..105S} have traced back the trajectories
of so-called 'field' O-stars and found that the overwhelming majority had their
origin in young open clusters. They found that the mass-loss rate from ejected O-stars
alone amounts to about 1.5 $M_\odot$ Myr$^{-1}$ during the first 6 Myr
of a cluster life. To get this number, a typical
O-star mass of 20 $M_\odot$ was assumed.

Provided that the cluster formation history has not \linebreak chang\-ed dramatically in the solar
neighbourhood during the evolution of the Galactic disk,
we estimate the contribution of mass from open clusters to the thin disk of the
Galaxy, or, to be more precise, the fraction of mass in the thin disk from stars that
have spent part of their lifetime being members of classical open clusters.

With an assumed lifetime of the thin disk of 10 Gyr, an average mass of open clusters
from the CIMF of $4500\,M_\odot$ and a cluster formation rate
of $0.4$ kpc$^{-2}$Myr$^{-1}$ we estimate this contribution to be
\[
\Sigma = 18\,M_\odot\,\textrm{pc}^{-2}\,.
\]
This has to be compared to the present total surface density {\em in form of stars} of the
Galactic disk in the solar neighbourhood that, according to \citet{2004MNRAS.352..440H}\
is $35\pm6\, M_\odot\,\textrm{pc}^{-2}$. As part of this mass is re-processed (mass-loss
from massive stars) one finds \citep{just09} that the amount of mass in stars ever formed 
in the thin disk must have been $48\pm6\, M_\odot\,\textrm{pc}^{-2}$ to explain the
present-day Holmberg \& Flynn value.
With these numbers about 37\% of the observed
surface density of the thin disk comes from open clusters.

This is considerably higher
than the previous estimates for the input of open clusters to the observed
stellar population of the Galactic disk that is quoted as about 10\%
\citep[see][]{1978PASP...90..506M,2006A&A...445..545P}
or even less than 10\% \citep{1971A&A....13..309W}.

\section{Summary and outlook}

Summing all this up, it is fair to say that this work on the population of open clusters
in the solar neighbourhood that started from an all-sky survey has found that
open clusters are larger, more massive, live longer and contribute more
to the thin stellar disk of the Galaxy than was believed a decade before.

Although much progress in our knowledge of the statistical properties
of the open cluster population has been made in the past decade,
there is still a long way to go. The sample discussed here allows us
to draw general
conclusions, but in some cases the derived parameters are erroneous
because of a severe undersampling of the cluster membership due to
the
bright magnitude limit of the \linebreak \ascc
survey.
New questions have appeared, such as:
is the local cluster population representative for the whole disk?
how to discern open clusters from compact associations? are these
separate populations
or can the latter be seen as the high-mass end of the cluster
population?
What exactly do we mean by ''the mass of a cluster''?
Tidal, virial and star-counted masses may not necessarily measure the
same mass.
To get more insight into this problem, one  must carefully define what
a ''cluster member'' is.
This becomes especially important near the boundary of a cluster and
consequently
influences our understanding of ''mass''.

Progress on some of these topics may be expected from an exploitation of the
new deep survey catalogue PPMXL \citep{2010arXiv1003.5852R}, and, on a
somewhat
longer timescale from Gaia.


\acknowledgements
The work on which this highligh talk is based upon was supported by
DFG grants 436 RUS 113/757/0-1 and 436 RUS 113/757/0-2,
and RFBR grants 03-02-04028, 06-02-16379 and 07-02-91566.


\end{document}